# Title: Giant and nonreciprocal second harmonic generation from layered antiferromagnetism in bilayer CrI$_3$

**Authors:** Zeyuan Sun[1†], Yangfan Yi[1†], Tiancheng Song[2], Genevieve Clark[3], Bevin Huang[2], Yuwei Shan[1], Shuang Wu[1], Di Huang[1], Chunlei Gao[1,4], Zhanghai Chen[1,4], Michael McGuire[5], Ting Cao[3,6], Di Xiao,[7] Wei-Tao Liu[1,4], Wang Yao[8], Xiaodong Xu[2,3*], Shiwei Wu[1,4*]

**Affiliations:**
[1] State Key Laboratory of Surface Physics, Key Laboratory of Micro and Nano Photonic Structures (MOE), and Department of Physics, Fudan University, Shanghai 200433, China.
[2] Department of Physics, University of Washington, Seattle, Washington 98195, USA
[3] Department of Materials Science and Engineering, University of Washington, Seattle, Washington 98195, USA
[4] Collaborative Innovation Center of Advanced Microstructures, Nanjing 210093, China.
[5] Materials Science and Technology Division, Oak Ridge National Laboratory, Oak Ridge, Tennessee, 37831, USA
[6] Geballe Laboratory for Advanced Materials, Stanford University, Stanford, California 94305, USA
[7] Department of Physics, Carnegie Mellon University, Pittsburgh, Pennsylvania 15213, USA
[8] Department of Physics and Center of Theoretical and Computational Physics, University of Hong Kong, Hong Kong, China

† These authors equally contributed to this work.
* Corresponding emails: swwu@fudan.edu.cn, xuxd@uw.edu

**Abstract**: Layered antiferromagnetism is the spatial arrangement of ferromagnetic layers with antiferromagnetic interlayer coupling. Recently, the van der Waals magnet, chromium triiodide (CrI$_3$), emerged as the first layered antiferromagnetic insulator in its few-layer form[1], opening up ample opportunities for novel device functionalities[2-7]. Here, we discovered an emergent nonreciprocal second order nonlinear optical effect in bilayer CrI$_3$. The observed second harmonic generation (SHG) is giant: several orders of magnitude larger than known magnetization induced SHG[8-11] and comparable to SHG in the best 2D nonlinear optical materials studied so far[12-15] (e.g. MoS$_2$). We showed that while the parent lattice of bilayer CrI$_3$ is centrosymmetric and thus does not contribute to the SHG signal, the observed nonreciprocal SHG originates purely from the layered antiferromagnetic order, which breaks both spatial inversion and time reversal symmetries. Furthermore, polarization-resolved measurements revealed the underlying $C_{2h}$ symmetry, and thus monoclinic stacking order in CrI$_3$ bilayers, providing crucial structural information for the microscopic origin of layered antiferromagnetism[16-20]. Our results highlight SHG as a highly sensitive probe that can reveal subtle magnetic order and open novel nonlinear and nonreciprocal optical device possibilities based on 2D magnets.

**Main text**:

Second harmonic generation (SHG) is a nonlinear optical process which converts two photons of the same frequency into a photon of twice the fundamental frequency. It is not only of technological importance for nonlinear optical devices, but also a powerful tool for the investigation of symmetry-related physical phenomena that are otherwise challenging to probe[21,22]. The power of this technique lies in its sensitivity to inversion symmetry breaking, the prerequisite for nonvanishing SHG under the electric dipole approximation. For a system without lattice inversion symmetry, SHG is electric dipole allowed and known as a time-invariant or i-type process. In the presence of lattice inversion symmetry, SHG can also be allowed if, for instance, there is an underlying magnetic structure that breaks both spatial inversion and time-reversal symmetries[8-11]. This time-noninvariant or nonreciprocal SHG, denoted as c-type, is also electric dipole allowed. Compared to i-type SHG, c-type is less common and often a weaker effect, which has been utilized to probe antiferromagnetic order in bulk crystals such as $Cr_2O_3$[9-11,23] and surface ferromagnetism in transition metal thin films[8,24,25].

The recent discovery of 2D van der Waals magnets[1,26-32] may provide a new platform for exploring second order nonlinear optical effects. Among these magnets, bilayer $CrI_3$ is particularly interesting due to the interplay between its crystal structure and magnetic order. As shown by Fig. 1a, monolayer $CrI_3$ has a centrosymmetric lattice structure with three-fold rotational symmetry[16]. When two monolayer sheets are stacked along the same orientation (Fig. 1b), the $CrI_3$ bilayer remains centrosymmetric, regardless of any rigid translation between the two sheets. Therefore, i-type SHG in $CrI_3$ bilayers is forbidden under the electric dipole approximation[21]. On the other hand, c-type SHG originating from the magnetic structure could arise due to the layered antiferromagnetic order[9]. As shown in Fig. 1c, the two layered antiferromagnetic configurations, with all spins pointing outward or inward, break both time reversal and spatial inversion symmetries, allowing electric dipole c-type SHG. In contrast, when the bilayer is driven into the fully spin aligned states upon the application of an out-of-plane magnetic field, the inversion symmetry of the magnetic structure is restored (Fig. 1d) and prohibits c-type SHG. These unique magnetic structures in $CrI_3$ bilayers that can be possibly tuned thus allow for the exploration of magnetization-induced electric dipole SHG in the atomically thin limit, which, in turn, may reveal subtle structural information of the magnetism that cannot be determined with existing approaches.

We first investigate the layered antiferromagnetism induced nonreciprocal SHG in $CrI_3$ bilayers as a function of temperature and out-of-plane magnetic field. Unless otherwise noted, a pulsed 900-nm femtosecond laser at a power of 0.6 mW was used (see details in Methods). $CrI_3$ bilayers were mechanically exfoliated and encapsulated by thin hexagonal boron nitride (hBN) flakes to prevent degradation[1]. Figure 1e shows an optical microscope image of the sample, along with a neighboring thicker flake, before hBN encapsulation from which the data in the main text were taken. Figures 1f-h show SHG microscope images of the same area as in Fig. 1e. At 50 K, above the Néel temperature (~40 K) of bilayer $CrI_3$, no SHG signal is observed (Fig. 1f). When the sample was cooled down to 5 K at zero magnetic field, a layered antiferromagnetic state forms, and a strong and homogenous SHG signal emerges (Fig. 1g). The signal from the bilayer vanishes at a magnetic field of -1 T (Fig. 1h) which aligns all spins. The isolated $CrI_3$ thicker flake has a slightly different SHG response compared to the bilayer due to its increased thickness and differing magnetic structure (Extended Data Fig. 1), which is not the focus of this work. The SHG signal from thin hBN flakes used for encapsulation is hardly observed since the bulk crystal of hBN is centrosymmetric[12].

We extract the SHG intensity of the bilayer at zero magnetic field and plot it as a function of temperature in Fig. 1i. The SHG signal depends sensitively on the temperature and vanishes above the Néel temperature where the antiferromagnetic order disappears. Similar behavior was observed on other CrI$_3$ bilayers, as shown in Extended Data Fig. 2. The vanishing SHG above the Néel temperature as well as in the fully spin aligned states also confirms the centrosymmetric lattice structure for bilayers of CrI$_3$. The combined magnetic field and temperature dependent measurements unambiguously point to layered antiferromagnetic order as the origin of the SHG.

While the measurements are consistent with the symmetry analysis above, the observation of such a strong c-type SHG is remarkable, because the magnetization-induced SHG was generally rather weak[8,9]. For example, in Cr$_2$O$_3$ bulk (a model system for studying antiferromagnetism-induced SHG), the electric dipole allowed c-type SHG is comparable to the magnetic dipole contribution of i-type SHG from the lattice[23]. The corresponding value of the c-type $|\chi^{(2)}|$ is only about 1-10 pm/V[33]. The electric dipole allowed c-type SHG was also detected at the surface or interface of ferromagnetic transition metal thin films, where both spatial inversion and time reversal symmetries are broken. The corresponding value of the c-type $|\chi^{(2)}|$ was found to be significantly weaker, on the order of $10^{-7}$ pm/V[24,25]. Negligible SHG in the CrI$_3$ bilayer with spin fully aligned states corroborates this weak surface effect.

To quantify the strength of the layered antiferromagnetism-induced SHG in bilayer CrI$_3$, we measured it along with monolayers of MoS$_2$, WSe$_2$, and hBN in the same experimental set-up (Fig. 2). Here, the MoS$_2$ and WSe$_2$ monolayers were exfoliated onto SiO$_2$/Si substrates, while the hBN monolayer was grown by chemical vapor deposition and transferred onto a SiO$_2$/Si substrate. Following the procedure detailed in Ref. 34, we deduce the second-order nonlinear susceptibilities $|\chi^{(2)}|$ of each of the aforementioned materials. With 900-nm excitation (i.e. 450-nm SHG emission), the $|\chi^{(2)}| \sim 2$ nm/V of CrI$_3$ bilayers is about 5 times stronger than that of hBN monolayers and is about the same order of magnitude as monolayers of MoS$_2$ and WSe$_2$ ($\left|\chi^{(2)}_{MoS_2}\right|:\left|\chi^{(2)}_{WSe_2}\right|:\left|\chi^{(2)}_{CrI_3}\right| \sim 4:3:1$). The magnitude of the bilayer CrI$_3$ SHG signal is also comparable to that of the MoS$_2$ monolayer SHG signal at the 1s exciton resonance (black arrow in Fig. 2). Knowing that monolayers of MoS$_2$ and WSe$_2$ exhibit the strongest i-type $|\chi^{(2)}|$ among all the existing 2D materials and are also comparable to the best-known second order nonlinear crystals[12-15], strong c-type SHG from the layered antiferromagnetism in bilayer CrI$_3$ is extraordinary.

This giant c-type SHG enables the probe of the structural symmetry of layered antiferromagnetic states in bilayer CrI$_3$, for which conventional means including neutron diffraction and the magneto-optical Kerr effect are hardly applicable[10]. Such information is currently lacking but crucial for understanding the microscopic origin of the antiferromagnetic interlayer coupling in bilayer CrI$_3$. In particular, while bulk CrI$_3$ crystals are known to be ferromagnetic[16], CrI$_3$ bilayers exhibit layered antiferromagnetism. Very recently, theoretical calculations[17-20] show that the magnetic interlayer coupling in CrI$_3$ depends critically on the stacking structure, i.e. how the two monolayer sheets in the bilayer are laterally translated with respect to each other. In particular, the rhombohedral stacking structure is predicted to favor ferromagnetic interlayer coupling, while the monoclinic stacking structure is antiferromagnetically coupled. On the other hand, it was reported that the bulk CrI$_3$ crystal has a rhombohedral structure at low temperature and a monoclinic structure at high temperature, with a structural phase transition occurring at around ~200 K[16]. Thus, the stacking structure, or magnetic structure due to spin-lattice coupling, in bilayer CrI$_3$ at low temperatures is an open question.

As illustrated in Extended Data Fig. 3, the rhombohedral structure belongs to the $S_6$ crystallographic point group which possesses an out-of-plane $C_3$ axis and lacks a mirror plane. In contrast, the monoclinic structure possesses $C_{2h}$ symmetry, which has an in-plane $C_2$ axis and a mirror plane. Polarization-resolved SHG measurements are ideally suited to distinguish the symmetry difference between the two stacking structures. Briefly, for a system with three-fold rotational symmetry, because of angular momentum conservation associated with the photon's helicity[35], the SHG process should have a cross-circularly polarized optical selection rule as depicted in Fig. 3a. That is, the absorption of two $\sigma^+$ photons at the fundamental frequency leads to the emission of one $\sigma^-$ photon at twice the fundamental, and vice versa, as demonstrated in monolayer transition metal dichalcogenides[36,37]. However, while our measurements at zero applied magnetic field show that cross-circularly polarized SHG is indeed the strongest component in the antiferromagnetic state (Figs. 3b-c), the SHG signals for co-circularly polarized excitation and detection are appreciable (Figs. 3d-e). Figure 3f shows the corresponding SHG spectra with co- and cross-polarized measurements. Such observations suggest the lack of three-fold rotational symmetry in its spin-lattice structure, i.e. CrI$_3$ bilayers do not have a rhombohedral stacking structure at low temperatures.

This was further confirmed by polarization-dependent azimuthal SHG measurements in the linear polarization basis. In these measurements, the excitation and detection are either co- (XX) or cross- (XY) linearly polarized while rotating together with respect to the sample plane. Figure 3g shows the polarization dependence under 900-nm excitation. Both XX (black dots) and XY (red dots) patterns show six asymmetric lobes, confirming the broken three-fold rotational symmetry. In fact, the XX and XY patterns are nicely fitted to the solid curves obtained from the c-type second order nonlinear tensors associated with $C_{2h}$ crystallographic symmetry, which corresponds to the monoclinic stacking structure (see details in Methods). Such observations and fits are robust at different fundamental wavelength excitations. Figures 3h and i show the azimuthal polarization-dependence of the SHG signal at 970-nm and 1040-nm excitation, respectively. Although details of the SHG patterns vary with the excitation wavelength, the revealed $C_{2h}$ symmetry is consistent with an in-plane $C_2$ axis pointing to the same direction (~55° with respect to the laboratory x coordinate). The details of the SHG polarization dependence is likely determined by the electronic states at the excitation energy[38], which reflects the combined spin, rotation, and inversion symmetries of the system. Our SHG study thus provides the first experimental evidence to support the monoclinic structure at low temperatures. Such a structure is consistent with theoretical predictions[17-20], which show layered antiferromagnetism as the ground state in CrI$_3$ bilayers.

We further studied the polarization-resolved SHG associated with different magnetic states. Figure 4a shows the SHG intensity as a function of magnetic field under $\sigma^+$ excitation and $\sigma^-$ detection ($\sigma^+/\sigma^-$). For comparison, Fig. 4b shows the corresponding magnetic field-dependent reflectance magneto circular dichroism (RMCD) measurement of the same bilayer. Consistent with previous reports, RMCD is only non-zero when the bilayer is in one of the fully spin aligned ferromagnetic states, and vanishes in the antiferromagnetic state. The SHG signal, in contrast, is non-zero in the layered antiferromagnetic states. Examining the data in Fig. 4a, the SHG signal is significantly larger when the magnetic field is swept upward from -1 T than when the field is swept downward from 1 T. If the polarization setting is switched to $\sigma^-/\sigma^+$, equivalent to applying a time reversal operation on the spin-lattice structure, the intensities are reversed (Extended Data Fig. 4). This distinct difference in the SHG signal between the two field sweep directions demonstrates the existence of two antiferromagnetic ground states, as schematically shown in Fig. 4c. Thus the two antiferromagnetic ground states can be independently studied, as shown in Fig. 3 and Extended Data Fig. 5.

In addition, SHG can also probe magnetic domain dynamics near the metamagnetic transition[8,9]. Figures 4d-h shows SHG microscope images at select magnetic fields. At -0.63 T, no SHG signal in the CrI$_3$ bilayer was observed, implying that the whole bilayer was in the fully spin aligned state. As the field was swept to -0.58 T, the lower half of the CrI$_3$ bilayer displayed SHG signal, corresponding to domain switching from the spin-aligned state to the antiferromagnetic state. As the field was continuously swept up to -0.52 T, the SHG signal in the upper half of the CrI$_3$ bilayer gradually appeared in spots as more domains switched from the spin-aligned state to the antiferromagnetic state. We note that the domain switching observed by MOKE microscopy[1] is subtly different as MOKE is sensitive to ferromagnetic states. With sensitivity to the antiferromagnetic states, SHG microscopy provides a powerful technique to image antiferromagnetic domain switching in CrI$_3$ bilayers. Our results therefore highlight the opportunities of applying SHG to investigate 2D antiferromagnetic order as well as exploring nonreciprocal nonlinear optics with possible control at the atomically thin limit[5-7,39].

## Methods

**Sample preparation.** The CrI$_3$ bilayer samples were prepared on Si substrates with 285 nm thick SiO$_2$ by mechanical exfoliation from bulk single crystals and encapsulated by hBN thin flakes (about 20 nm thick). The whole process was done in a glove box with an argon atmosphere. The number of layers of the samples was determined by optical contrast as well as from the RMCD and photoluminescence measurements performed under an externally applied magnetic field.

**Optical measurements.** The experiments were conducted in a variable temperature optical cryostat housed inside a superconducting magnet with a room temperature bore in Faraday geometry. The RMCD measurement was based on the photoelastic modulation (PEM) technique. The wavelength was at 632.8 nm and the excitation power was 5 μW. The SHG measurement was conducted by using the femtosecond laser pulses from a Ti:Sapphire oscillator (MaiTai HP, Spectra Physics). The wavelength was tunable from 700 nm to 1040 nm, but a wavelength of 900 nm was chosen for most of the measurements. The laser pulses were focused onto the sample at normal incidence using a microscopic objective. The reflected SHG signal was collected by the same objective and detected by either a photomultiplier tube in photon counting mode or a spectrograph equipped with a liquid nitrogen-cooled CCD. The SHG microscope images were obtained by raster scanning the laser beam against the sample using a two-axis galvanometer. The polarized SHG images and azimuthal SHG polarization dependence were obtained by setting the excitation and detection beams with a quarter or half wave plate in combination with linear polarizers. Details of the optical layout are shown in Extended Data Fig. 6. Given the small sample size, just a few microns in length, the data points in the polarization dependence and magnetic hysteresis loops were obtained from a series of polarization-resolved SHG images at different azimuthal angles. To save time in the polarization dependence study, we measured only up to 180°, and projected the data at an azimuthal angle, θ, to that at θ + 180°. The legitimate use of this protocol was checked and passed by rotating the azimuthal angle in a complete 360° manner.

**SHG Pattern fitting.** The polarization-dependent patterns of the SHG including co-(XX) and cross-(XY) linear polarization configurations are fit together by considering the monoclinic stacking structure in a CrI$_3$ bilayer, which has the crystallographic symmetry $C_{2h} =$ $\{e, C_2, i, \sigma_\perp\}$, where $e$ is the identity operator, $C_2$ is a two-fold rotation operator with an in-plane axis, $i$ is the spatial inversion operator and $\sigma_\perp$ is the reflection operator whose mirror plane is normal to $C_2$. Taking the out-of-plane layered antiferromagnetic spin configuration into account, the corresponding magnetic point group is $C_{2h}(C_2) = \{e, C_2, Ri, R\sigma_\perp\}$, where $R$ is the time reversal operator. Let $x$ be the in-plane $C_2$ axis, $z$ be the out-of-plane direction and $yz$ be the mirror plane (Extended Data Fig. 3). Since the light propagation is along the $z$ axis (normal incidence), we only need to consider the $x$ and $y$ component of the second order susceptibility tensor $\chi_{ijk}^{(c)}$. Under $C_{2h}(C_2)$, there are three independent nonzero tensor elements: $\chi_{xxx}, \chi_{xyy}, \chi_{yxy} = \chi_{yyx}$. Let the azimuthal angle between the laboratory and crystallographic coordinates be $\varphi$. Then the measured SHG signals under XX and XY polarization configurations, respectively, take the following forms:

$$I_\parallel(\text{XX}) \propto [\chi_{xxx}\cos^3\varphi + (\chi_{xyy} + 2\chi_{yxy})\sin^2\varphi \cos\varphi]^2,$$
$$I_\perp(\text{XY}) \propto [\chi_{xyy}\sin^3\varphi + (\chi_{xxx} - 2\chi_{yxy})\sin\varphi \cos^2\varphi]^2.$$

Since the excitation and emission wavelengths of SHG involve the resonant electronic transitions in CrI$_3$, the three independent tensor elements are complex numbers in the fits.

**Acknowledgements:** We would like to thank Prof. Yuen-Ron Shen for insightful discussion. The work at Fudan was supported by the National Natural Science Foundation of China (11427902), National Basic Research Program of China (2014CB921601) and National Key Research and Development Program of China (2016YFA0301002). The work at the University of Washington and Carnegie Mellon University was mainly supported by the Department of Energy, Basic Energy Sciences, Materials Sciences and Engineering Division (DE-SC0012509). Device fabrication is partially supported by NSF-DMR-1708419. XX acknowledges the support from the State of Washington funded Clean Energy Institute and from the Boeing Distinguished Professorship in Physics.

**Author Contributions:** S.W. and X.X. conceived and supervised the project. Z.S., Y.Y., Y.S. and S.W. (Shuang Wu) conducted the measurements with technical assistance from B.H., D.H., C.G., Z.C. The samples were prepared by T.S. and G.C., and single crystals were supplied by M.M. Z.S., Y.Y., T.C., D.X., W.L., W.Y., X.X. and S.W. analyzed the data. Z.S., Y.Y., X.X. and S.W. wrote the paper with contributions from all authors.

**Competing Interests:** The authors declare no competing financial interests.

**Data Availability:** The datasets generated during and/or analyzed during this study are available from the corresponding author upon reasonable request.

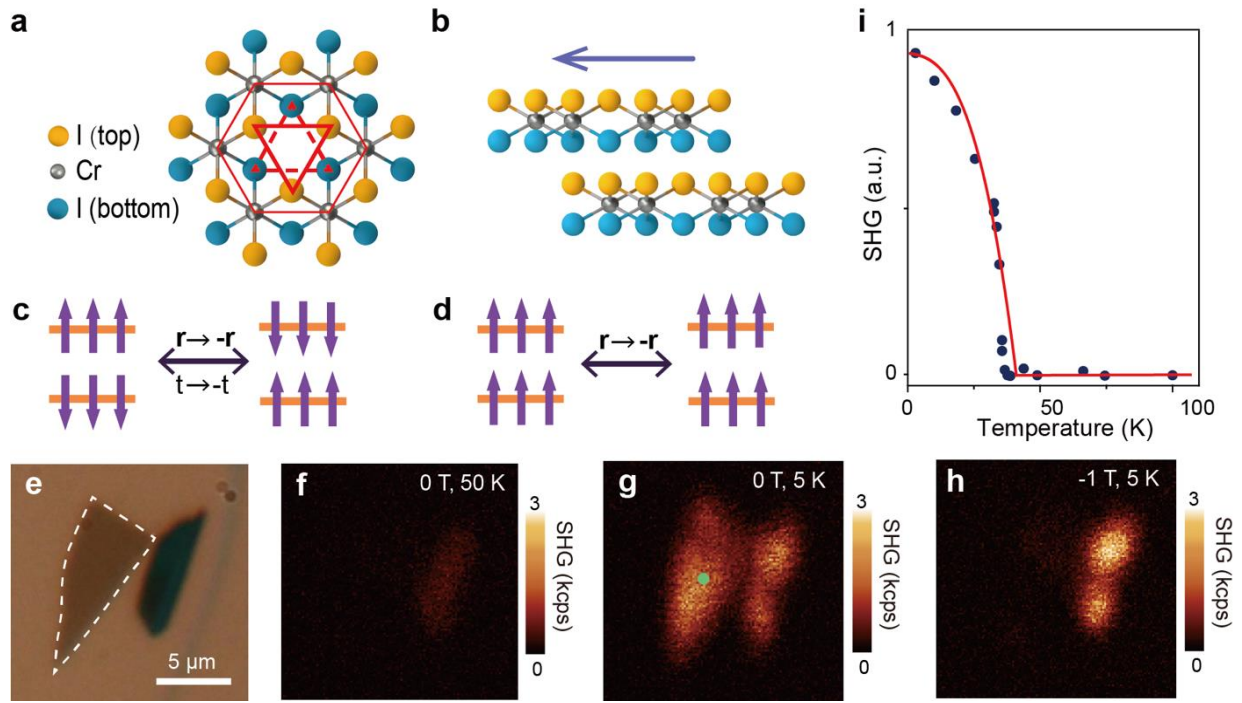

**Figure 1 | Second harmonic generation (SHG) response from a CrI$_3$ bilayer. a,** Atomic structure of CrI$_3$ monolayer. Six chromium atoms (gray) are connected by thin red solid lines forming a hexagon, three iodine atoms in the top (orange) and bottom (blue) layers are connected by solid and dashed red lines, respectively, forming equilateral triangles. **b**, Side view of a CrI$_3$ bilayer. The two layers are orientated in the same crystallographic direction with a possible lateral translation indicated by the arrow. **c,** Schematics of two antiferromagnetic states of a CrI$_3$ bilayer. Either a spatial inversion (**r** $\rightarrow$ -**r**) operation or a time reversal (t $\rightarrow$ -t) operation converts one antiferromagnetic state to the other, but not to itself. **d**, Schematic of a ferromagnetic state of a CrI$_3$ bilayer with the centrosymmetric spin-lattice structure. **e,** Optical microscope image of a CrI$_3$ bilayer (delineated by the white dotted line) and a thicker flake. Scale bar: 5 µm. **f-h**, The corresponding SHG intensity images when the bilayer is **f**, nonmagnetic (0 T, 50 K), **g**, antiferromagnetic (0 T, 5 K) and **h**, ferromagnetic (-1 T, 5 K). **i**, SHG intensity of the bilayer as a function of temperature. All data were taken at the green dot in **g**. The red solid curve is a guide-to-the-eye.

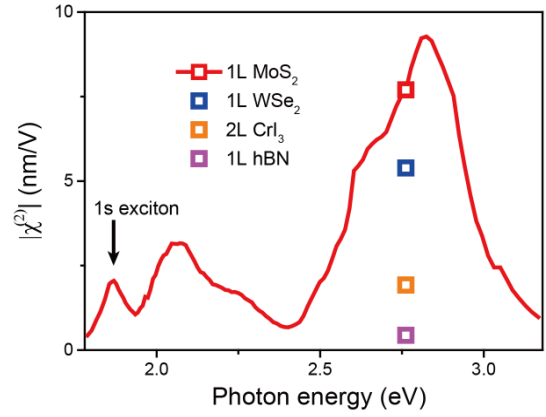

**Figure 2 | Comparison of the second-order nonlinear susceptibility in different 2D materials.** The open squares show the measured second-order susceptibilities $|\chi^{(2)}|$ of monolayer $MoS_2$ (red), monolayer $WSe_2$ (blue), monolayer hBN (purple) and bilayer $CrI_3$ in the antiferromagnetic state (orange). The measurement was conducted using the same set-up (Extended Data Fig. 6) with a second harmonic photon energy of 2.76 eV (450 nm). The red curve shows the $|\chi^{(2)}|$ spectrum of $MoS_2$ monolayer with the second harmonic photon energy ranging from 1.8 eV to 3.2 eV.

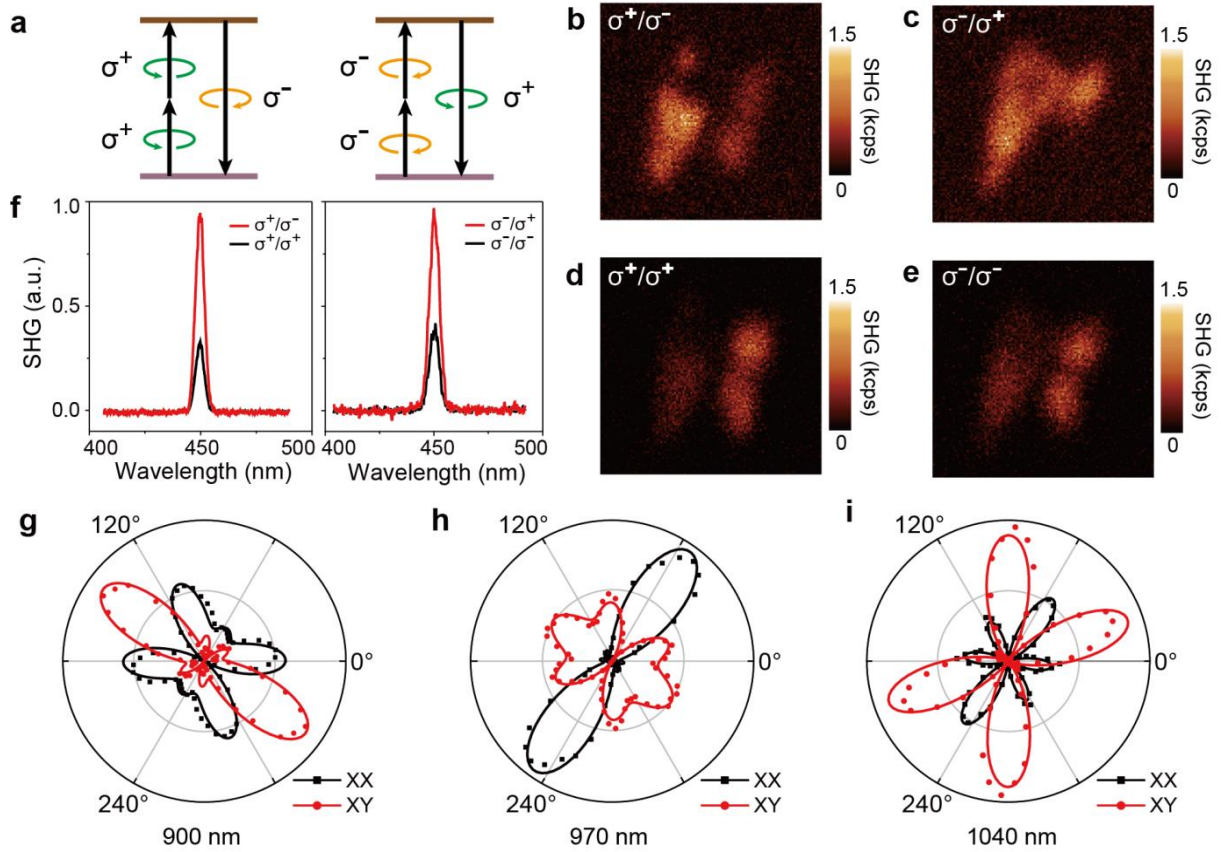

**Figure 3 | Polarization-resolved SHG in the layered antiferromagnetic state. a**, Circularly polarized SHG optical selection rules with three-fold rotational symmetry. The upward and downward arrows represent the fundamental and second-harmonic light, respectively. **b-e**, Polarization-resolved SHG intensity images at zero magnetic field: **b** $\sigma^+/\sigma^-$, **c** $\sigma^-/\sigma^+$, **d** $\sigma^+/\sigma^+$, and **e** $\sigma^-/\sigma^-$. **f**, Polarization-resolved SHG spectra. **g-i**, Azimuthal SHG polarization dependence at 0 T with the fundamental wavelength of **g**, 900 nm, **h**, 970 nm and **i**, 1040 nm. The excitation and detection beams were linearly polarized, with XX and XY referring to the two beams being co- and cross-linearly polarized, respectively. Data in (**f**) to (**i**) were obtained at the position marked by the green dot in Fig. 1g. Solid lines are fits by the c-type second order nonlinear tensors associated with $C_{2h}$ symmetry (monoclinic stacking structure), as described in Methods.

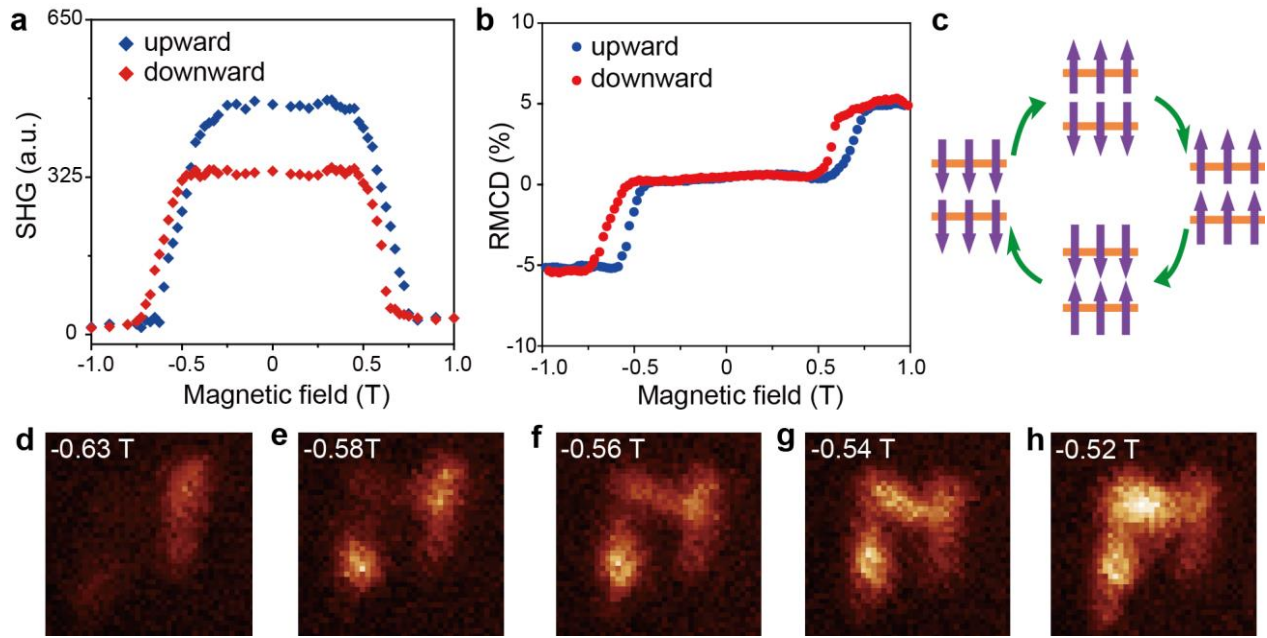

**Figure 4 | Probing the magnetic switching and domains. a**, Circularly polarized SHG intensity as a function of magnetic field. The excitation is σ⁺ polarized and the detection is σ⁻ polarized. The data were taken at the sample position marked by the green dot in Fig. 1g. **b**, Corresponding RMCD hysteresis loop at the same position in **a**. **c**, Schematic of the evolution of magnetic states driven by magnetic field. **d-h**, Circularly polarized SHG intensity plot with σ⁺/σ⁻ configuration at select magnetic fields near the metamagnetic transition while the field was swept up, revealing domain effects in the layered antiferromagnetic states.

**Extended Data for**

**Giant and nonreciprocal second harmonic generation from layered antiferromagnetism in bilayer CrI$_3$**

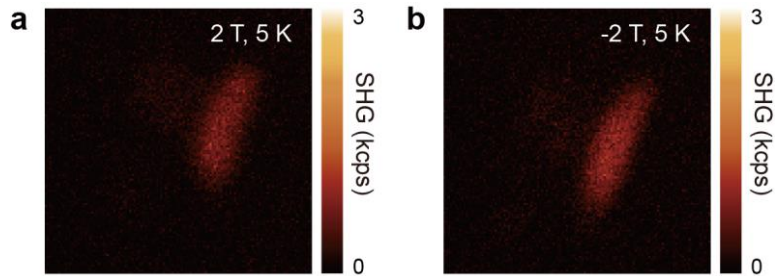

**Extended Data Figure 1 | SHG intensity images at higher magnetic field (±2 T).** Compared to the SHG intensity images in Fig. 1, no SHG was observed on the CrI$_3$ bilayer (left in the image) and the SHG becomes weaker on the thicker flake (right in the image).

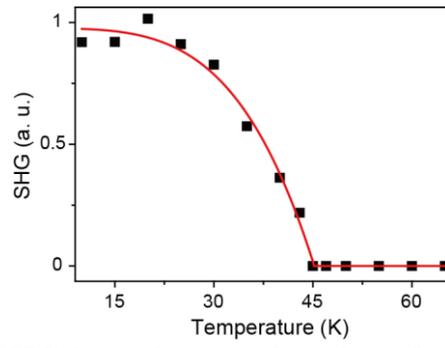

**Extended Data Figure 2 | SHG intensity as a function of temperature on another CrI$_3$ bilayer sample.** The bilayer was in a layered antiferromagnetic state with no external magnetic field applied. The red solid curve is a guide-to-the-eye, showing the transition temperature near 45 K.

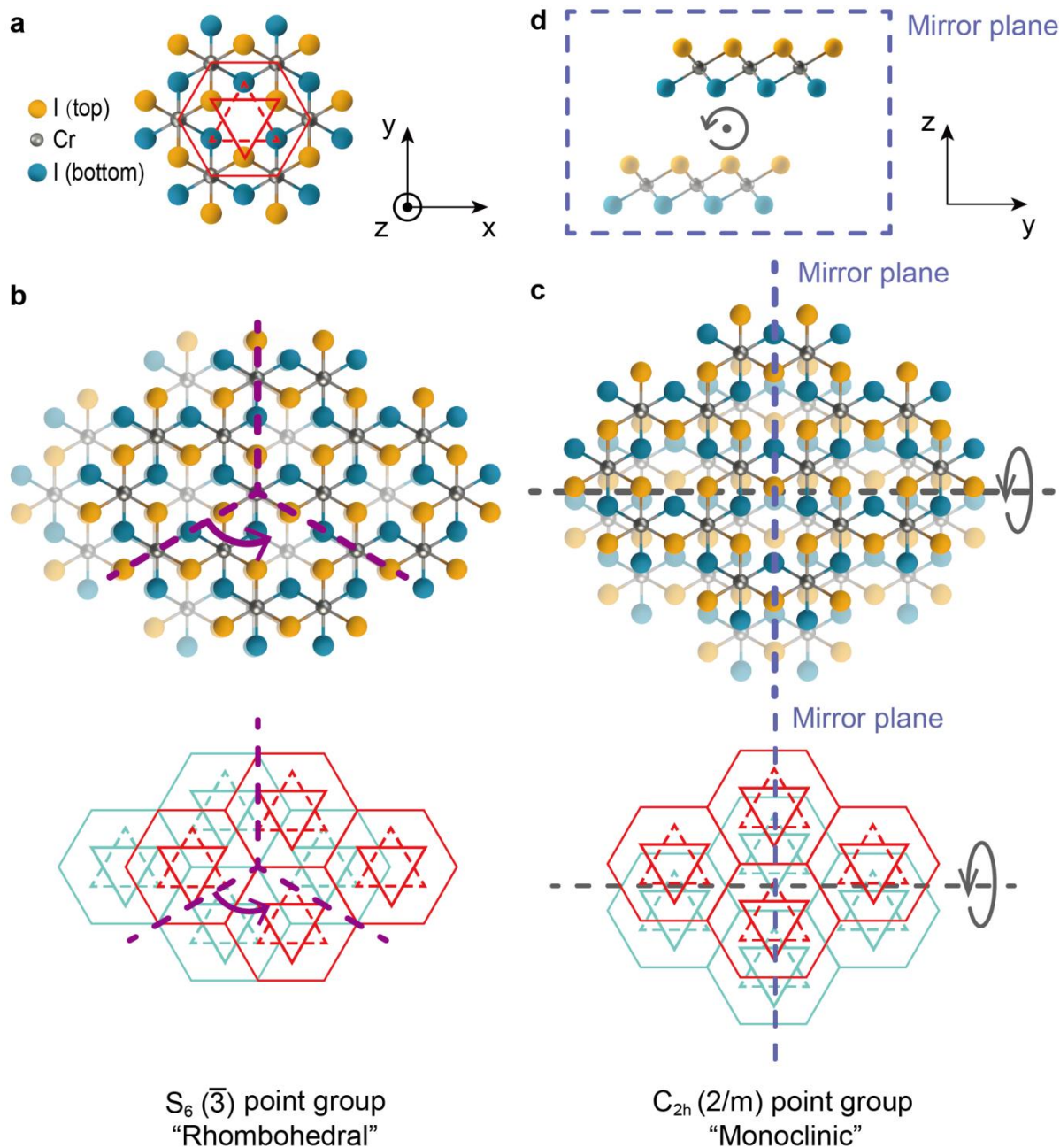

**Extended Data Figure 3 | Possible stacking structures in CrI$_3$ bilayer with distinct crystallographic symmetry. a,** Atomic structure of CrI$_3$ monolayer as in Fig. 1a. **b, c,** Rhombohedral and monoclinic stacking structures in bilayer CrI$_3$, respectively. The rhombohedral structure belongs to the $S_6$ crystallographic point group which has an out-of-plane $C_3$ axis and lacks a mirror plane. In contrast, the monoclinic structure possesses $C_{2h}$ symmetry which has an in-plane $C_2$ axis and a mirror plane. Note that if the monolayer sheets are laterally translated along the mirror plane in the monoclinic stacking structure, $C_{2h}$ symmetry remains. **d,** Side view of the monoclinic stacking structure.

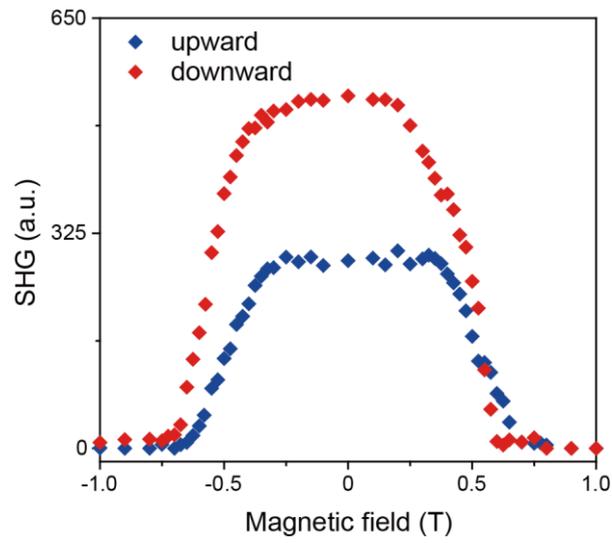

**Extended Data Figure 4 | Circularly polarized SHG intensity as a function of magnetic field.** The excitation is σ⁻ polarized and the detection is σ⁺ polarized. The data were taken at the sample position marked by the green dot in Fig. 1g.

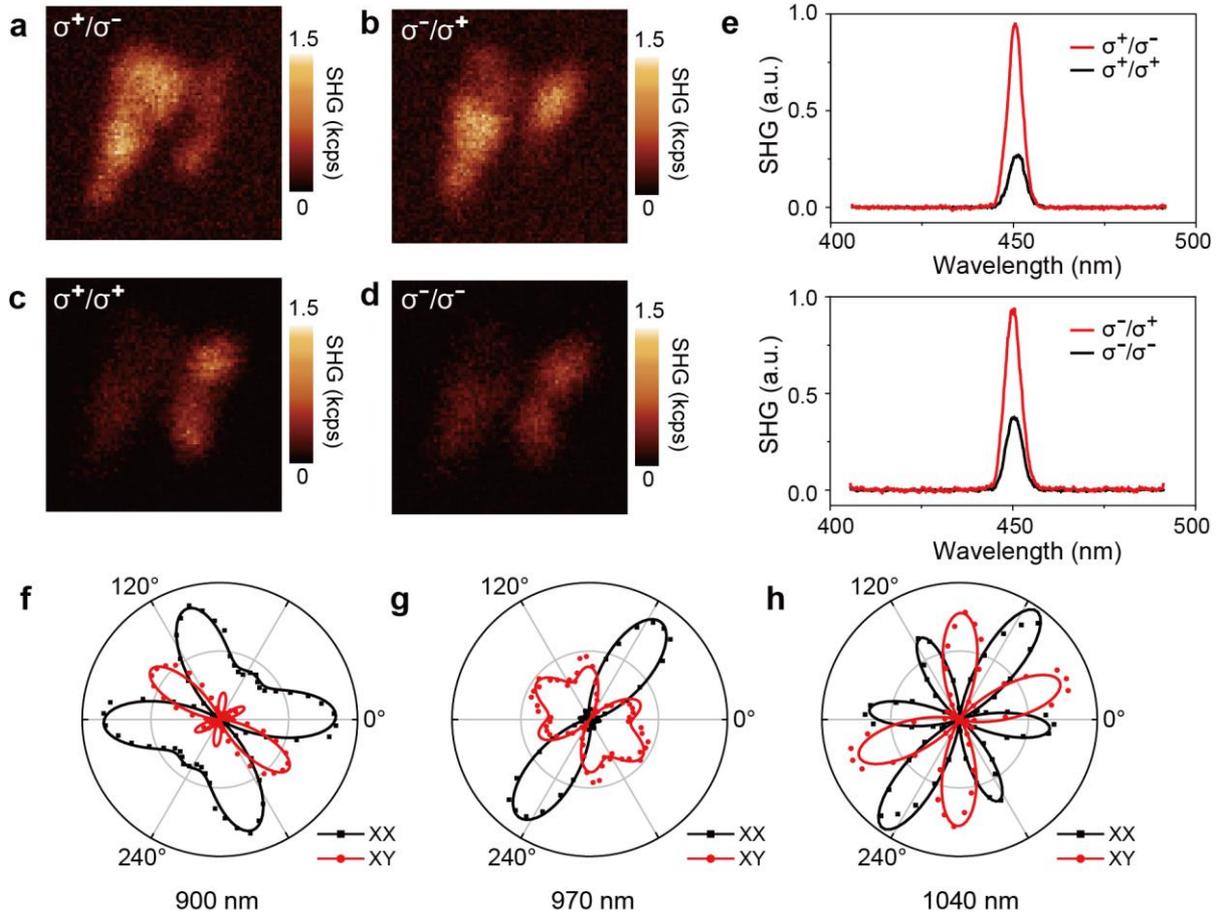

**Extended Data Figure 5 | Polarization resolved SHG in the other antiferromagnetic state. a-d**, Polarization resolved SHG intensity images at zero magnetic field: **a** $\sigma^+/\sigma^-$, **b** $\sigma^-/\sigma^+$, **c** $\sigma^+/\sigma^+$, **d** $\sigma^-/\sigma^-$. Here the magnetic field was swept upward from -1 T to 0 T, in contrast to the downward sweeping direction for Fig. 3. **e**, Corresponding polarization resolved SHG spectra. **f-h**, Azimuthal SHG polarization dependence at 0 T with the fundamental wavelength of **f**, 900 nm, **g**, 970 nm and **h**, 1040 nm. The excitation and detection beams were linearly polarized, with XX and XY referring to co- and cross-linearly polarized between the two beams, respectively. Data in (**e**) to (**h**) were obtained at the position marked by the green dot in Fig. 1g. Solid lines are fits by the c-type second order nonlinear tensors associated with $C_{2h}$ symmetry (monoclinic stacking structure), as described in Methods.

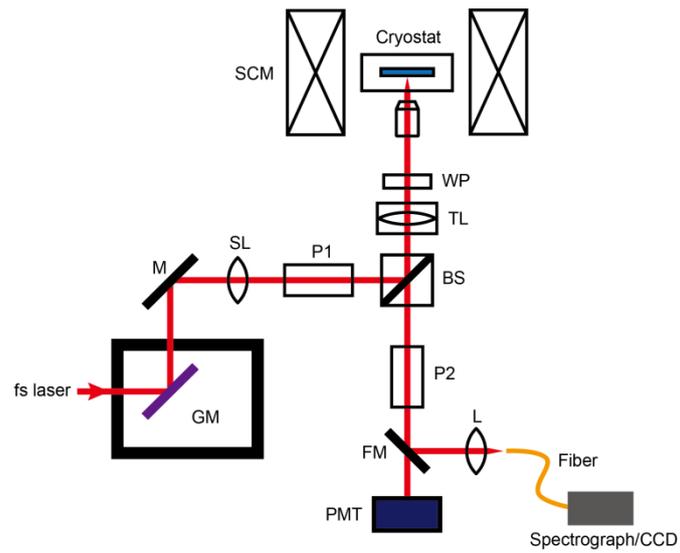

**Extended Data Figure 6 | Optical layout for the SHG measurement.** GM: two-axis galvanometer, M: silver mirror, SL: scan lens, P1: Glan-Thompson polarizer, BS: beamsplitter, TL: tube lens, WP: wave plate, SCM: superconducting magnet, P2: Glan-Thompson polarizer, FM: flip mirror, PMT: photomultiplier tube in photon counting mode, and L: focusing lens.